\documentclass[a4paper,fleqn,usenatbib]{mnras}

\usepackage{newtxtext,newtxmath}
\usepackage{threeparttable}
\usepackage[T1]{fontenc}
\usepackage{ae,aecompl}
\usepackage{threeparttable}

\usepackage{amssymb}	
\usepackage{graphics}
\usepackage{graphicx}	
\usepackage{color,soul}

\def\apj{ApJ}
\def\apjs{ApJS}
\def\mnras{MNRAS}
\def\nat{Nat}
\def\apjl{ApJ}
\def\aap{A\&A}
\def\aj{AJ}
\def\sovast{SvA}

\def\asec{\ifmmode ^{\prime\prime}\else$^{\prime\prime}$\fi}

\def\msun{M$_{\odot}$}

\def\it{\sl}
\def\degs{\ifmmode ^{\circ}\else$^{\circ}$\fi}
\def\amin{\ifmmode ^{\prime}\else$^{\prime}$\fi}
\def\asec{\ifmmode ^{\prime\prime}\else$^{\prime\prime}$\fi}
\def\fss{\hbox{$.\!\!^{\rm s}$}}        
\def\farcs{\hbox{$.\!\!^{\prime\prime}$}}  
\def\h{$^{\rm h}$}
\def\m{$^{\rm m}$}

\def\j0740{J0740+6620}
\def\psr{J0740}

\def\msun{M$_\odot$}
\def\fermi{\textit{Fermi}}
\def\degs{\ifmmode ^{\circ}\else$^{\circ}$\fi}
\def\amin{\ifmmode ^{\prime}\else$^{\prime}$\fi}

\def\eqalign#1{\null\,\vcenter{\openup1\jot \m@th
   \ialign{\strut\hfil$\displaystyle{##}$&$\displaystyle{{}##}$\hfil
   \crcr#1\crcr}}\,}

\title[The ultracool WD companion of PSR \j0740]{The ultracool helium-atmosphere white dwarf companion of PSR \j0740?}

\author[D. M. Beronya et al.]{D. M. Beronya$^{1}$\thanks{E-mail: daria.beronya@gmail.com},
A. V. Karpova$^{1}$,
A. Yu. Kirichenko$^{2,1}$, 
S. V. Zharikov$^{2}$, \newauthor 
D. A. Zyuzin$^{1}$, 
Yu. A. Shibanov$^{1,3}$
and
A. Cabrera-Lavers$^{4,5}$\\
$^{1}$Ioffe Institute, Politekhnicheskaya 26, St. Petersburg, 194021,  Russia \\
$^{2}$Instituto de Astronom\'ia, Universidad Nacional Aut\'onoma de M\'exico, Apdo. Postal 877, Baja California, M\'exico, 22800\\
$^{3}$Peter the Great St.~Petersburg Polytechnic University, Politekhnicheskaya 29, St. Petersburg, 195251, Russia \\
$^{4}$Instituto de Astrof\'isica de Canarias, La Laguna 38200, Tenerife, Spain\\
$^{5}$Departamento de Astrof\'isica, Universidad de La Laguna, La Laguna 38206, Tenerife, Spain
}

\date{Accepted XXX. Received YYY; in original form ZZZ}

\pubyear{2019}

\begin{document}
\label{firstpage}
\pagerange{\pageref{firstpage}--\pageref{lastpage}}
\maketitle

\begin{abstract}
We report detection of the likely companion of the binary millisecond pulsar \j0740\ 
with the Gran Telescopio Canarias in the $r'$ and $i'$ bands. 
The position of the detected starlike source coincides with the pulsar coordinates 
within the $1\sigma$ uncertainty of $\approx$ 0.2 arcsec. 
Its magnitudes are $r'=26.51\pm0.17$ and $i'=25.49\pm0.15$.   
Comparing the data with the white dwarf cooling tracks suggests that 
it can be an ultracool helium-atmosphere white dwarf with 
the temperature $\lesssim3500$ K and cooling age $\gtrsim5$ Gyr. 
The age is consistent with the pulsar characteristic age corrected for kinematic effects. 
This is the reddest source among known white dwarf companions
of millisecond pulsars. Detection of the source in other bands 
would be useful to clarify its properties and nature. 
\end{abstract}

\begin{keywords}
binaries: general -- pulsars: individual: PSR \j0740
\end{keywords}


\section{Introduction}

Millisecond pulsars (MSPs) are neutron stars (NSs) with short rotation periods $P$, typically below 30 ms. 
They have small spin-down rates ($\dot{P}\lesssim10^{-19}$ s~s$^{-1}$) 
and constitute a special subclass on the $P-\dot{P}$ diagram, 
distinct from the population of `normal' pulsars \citep[e.g.,][]{manchester2017}. 
The first MSP, PSR B1937+21, was discovered in 1982 \citep{backer1982}. 
It is believed that these objects are formed in binary systems: 
the initially more massive of the two binary components star turns into a NS 
which is then spun-up (or `recycled') due to accretion of matter 
from the secondary star \citep{Bisnovatyi-Kogan1974,alpar1982,tauris2011}. 
This is supported by the fact that a high ($>50$ per cent) fraction of MSPs 
has companions and by discoveries of several MSPs switching between the rotation-powered 
and accretion-powered pulsar states \citep{Archibald2009,Papitto2013,Bassa2014,Roy2015}. 
Moreover, the NS binary 1FGL J1417.7--4402 was recently found to be in the late stages 
of the canonical MSP recycling process \citep[see][and references therein]{swihart2018}.  

Spin periods of MSPs are extraordinarily stable which may be related 
to their weak external magnetic fields \citep{manchester2017}. 
The period stability makes MSPs important in different areas of astrophysics. 
For instance, studies of the binary pulsar PSR B1913+16, 
the double-pulsar system PSR J0737$-$3039A/B 
and especially the triple system PSR J0337+1715
allow one to carry out unprecedented tests of general relativity 
\citep[][and references therein]{kramer2018,archibald2018,weisberg2016}. 
A set of MSPs (or `Pulsar Timing Array') was proposed for detection of 
low-frequency gravitational waves and spacecraft navigation \citep{manchester2017b,becker2018}. 
MSP timing is also used to obtain physical characteristics of stellar clusters \citep{freire2017,prager2017}. 
Studies of MSPs are essential for exploration of binary and stellar evolution 
and also reveal properties of the interstellar medium (ISM). 
Binary systems containing a MSP allow for most accurate determination 
of NSs masses which plays a key role in constraints on equation-of-state (EoS)
of the superdense matter in NSs interiors \citep{lattimer2012,antoniadis2013}. 

In a binary system masses of both the NS and its companion 
can be inferred with a high precision through the measurement 
of the general relativistic Shapiro delay
from pulsar radio timing observations \citep{shapiro1964,jacoby2005}. 
However, the effect is strong for edge-on systems;
otherwise, additional optical observations are necessary. 
The latter are also crucial for determination of an MSP companion nature  
\citep[e.g.][]{vankerkwijk2005}. 
Low-mass white dwarfs (WDs) are the most frequent MSP companions \citep{tauris2012}. 
 In such a case, comparison between the optical data and WD evolutionary sequences
can provide the type, temperature, cooling age and mass of the WD. 
Knowing the companion mass, the mass function (from pulsar timing observations)
and the mass ratio (from the optical phase-resolved spectroscopy),
one can estimate 
the pulsar mass and the system inclination. Unfortunately, due to the weakness of the optical counterparts, 
identifications of MSP companions are not numerous. 
Nevertheless, 
increasing the number of such identifications is 
important for understanding the nature and origin of these systems. 
The situation has been improving thanks to the world's largest telescopes 
and deep sky surveys \citep[e.g.,][]{bassa2016,dai2017,karpova2018,kirichenko2018}.

\begin{table}
\caption{Parameters of the \psr\ system 
(from \citet{arzoumanian2018}). 
Current values are available in the NANOGrav Data Repository$^a$. 
The dispersion measure distances $D_{\rm YMW}$ and $D_{\rm NE2001}$ were estimated
using the YMW16 \citep*{ymw} and NE2001 \citep{ne2001} models for the distribution 
of free electrons in the Galaxy, respectively. 
The distance $D_p$ corresponds to the timing parallax of 2.3(6) mas. 
Numbers in parentheses are 1$\sigma$ uncertainties related to the last significant digits quoted.} 
\begin{tabular}{lc}
\hline
Right ascension $\alpha$ (J2000)                                         & 07\h40\m45\fss79492(2) \\
Declination $\delta$(J2000)                                              & +66\degs20\amin33\farcs5593(2)\\
Galactic longitude $l$ (deg)                                             & 149.7 \\
Galactic latitude $b$ (deg)                                              & 29.6 \\
Epoch (MJD)                                                              & 57017 \\
Proper motion $\mu_\alpha =\dot{\alpha}{\rm cos}\delta$ (mas yr$^{-1}$)  & $-$10.3(2) \\
Proper motion $\mu_\delta$ (mas yr$^{-1}$)                               & $-$31.0(2) \\
Spin period $P$ (ms)                                                     & 2.88573641085019(2)\\
Period derivative $\dot{P}$ ($10^{-20}$ s s$^{-1}$)                      & 1.2184(4) \\
Characteristic age $\tau$ (Gyr)                                           & 3.75\\
Spin-down luminosity $\dot{E}$ (erg s$^{-1}$)                            & $2.0\times10^{34}$ \\
\hline
Orbital period $P_b$ (days)                                              & 4.766944619(1) \\
Projected semi-major axis (lt-s)                                         & 3.9775602(2) \\
Eccentricity (10$^{-6}$)                                                 & 4.9(1) \\
\hline
 Mass function $f_{\rm M}$ (\msun)$^b$                          & 0.003 \\
 Minimum companion mass $M_{c, \rm min}$ (\msun)$^b$            & 0.2 \\
\hline
Dispersion measure (DM, pc cm$^{-3}$)                                    & 15.0 \\
Distance $D_{\rm YMW}$ (kpc)                                             & 0.93 \\
Distance $D_{\rm NE2001}$ (kpc)                                          & 0.68 \\
Distance $D_{p}$ (kpc)                                                   & 0.4$^{+0.2}_{-0.1}$ \\
\hline 
\end{tabular}
\label{tab:param}
\begin{tablenotes}
\item $^a$https://data.nanograv.org/
\item  $^b$Parameters are provided by \citet{lynch2018}. 
$M_{c,\rm min}$ is calculated assuming the inclination angle $i=90$ deg 
and the pulsar mass $M_p=1.4$ \msun.
\end{tablenotes}
\end{table}

The subject of this paper, the binary MSP \j0740\ (hereafter \psr), was discovered in the radio 
in the frame of the Green Bank North Celestial Cap Pulsar Survey \citep{stovall2014,lynch2018}. 
The pulsar was also identified in gamma-rays with the  
\fermi\ 
telescope \citep{laffon2015, guillemot2016}. 
Table~\ref{tab:param} summarises the parameters of the pulsar system. 
Basing on the computed minimum companion mass of $\approx0.2$ \msun, 
\citet{lynch2018} suggested that the pulsar companion is a He WD. 
Using the magnitude lower limits from the PanSTARRS 3$\pi$ Steradian Survey bands, 
they  constrained the WD temperature $<4200$~K and age $>3.2$ Gyr. 

To find the optical counterpart of the pulsar companion, we performed deep optical observations 
with the Gran Telescopio Canarias (GTC). 
Here we report a likely identification of the pulsar companion. 
The details of the observations and data reduction are described in Sect.~\ref{sec:data}, 
the analysis and results are presented in Sect.~\ref{sec:results} and discussed in Sect.~\ref{sec:discussion}.

\section{Observations and data reduction} 
\label{sec:data}

The observations of the \psr\ field were performed on December 26, 
2017\footnote{Proposal GTCMULTIPLE2A-17BMEX, PI A. Kirichenko} in the Sloan $r'$ and $i'$ bands using 
the GTC/OSIRIS\footnote{{http://www.gtc.iac.es/instruments/osiris/}} instrument. 
The conditions were photometric and seeing varied from 0.8 to 1.3 arcsec. 
The instrument field of view (FOV) was 7.8 arcmin $\times$ 7.8 arcmin  with the pixel scale of $0.254$ arcsec. 
The pulsar was exposed on the CCD2 chip of the two OSIRIS CCD chips. 
The dithered science frames of $120$ s were taken, resulting in the total exposure times 
of $3.7$ and $2.4$~ks for the $r'$ and $i'$ filters, respectively. 
The short 15 s exposures were obtained in each band to avoid saturation 
of bright stars that were further used for precise astrometry. 
The observing log is presented in Table~\ref{tab:gtc}. 

  \begin{table}
      \centering
      \caption{Log of the GTC observations. The mean airmass and seeing values are presented.} 
      \label{tab:gtc}
      \begin{tabular}{cccc}
      \hline
      Filter & Exposure time, s & Airmass & Seeing, arcsec \\
      \hline
      $r'$ & $15\times1$   & 1.32 & 0.9 \\
           & $120\times31$ & 1.29 & 0.9 \\
      \hline
      $i'$ & $15\times1$   & 1.26 & 0.8 \\
           & $120\times20$ & 1.26 & 0.8 \\
      \hline
      \end{tabular}
  \end{table}

The data reduction and analysis were carried out within the Image Reduction 
and Analysis Facility ({\sc iraf}) package.  
  The raw data included 10 bias frames and 5 sky flats for each filter, which were used to create  
the master bias and master flats frames. 
The bias subtracted and the flat-fielded   science frames were obtained with the {\sc ccdproc} routine, aligned to the  frame  with the best seeing,  and then stacked for each filter using 
the {\sc imcombine} task. 
Accounting for known OSIRIS geometrical distortions and background gradients in the images towards 
the FOV boundaries\footnote{http://www.gtc.iac.es/instruments/osiris/media/OSIRIS-USER-MANUAL\_v3\_1.pdf}, 
we ignored the areas outside 2.5 arcmin from the target to avoid the influence of these effects 
on the calibration and further results.

\begin{figure*}
\setlength{\unitlength}{1mm}
\resizebox{15.cm}{!}{
\begin{picture}(130,63)(0,0)
\put(-15,0){ \includegraphics[width=8cm,clip=]{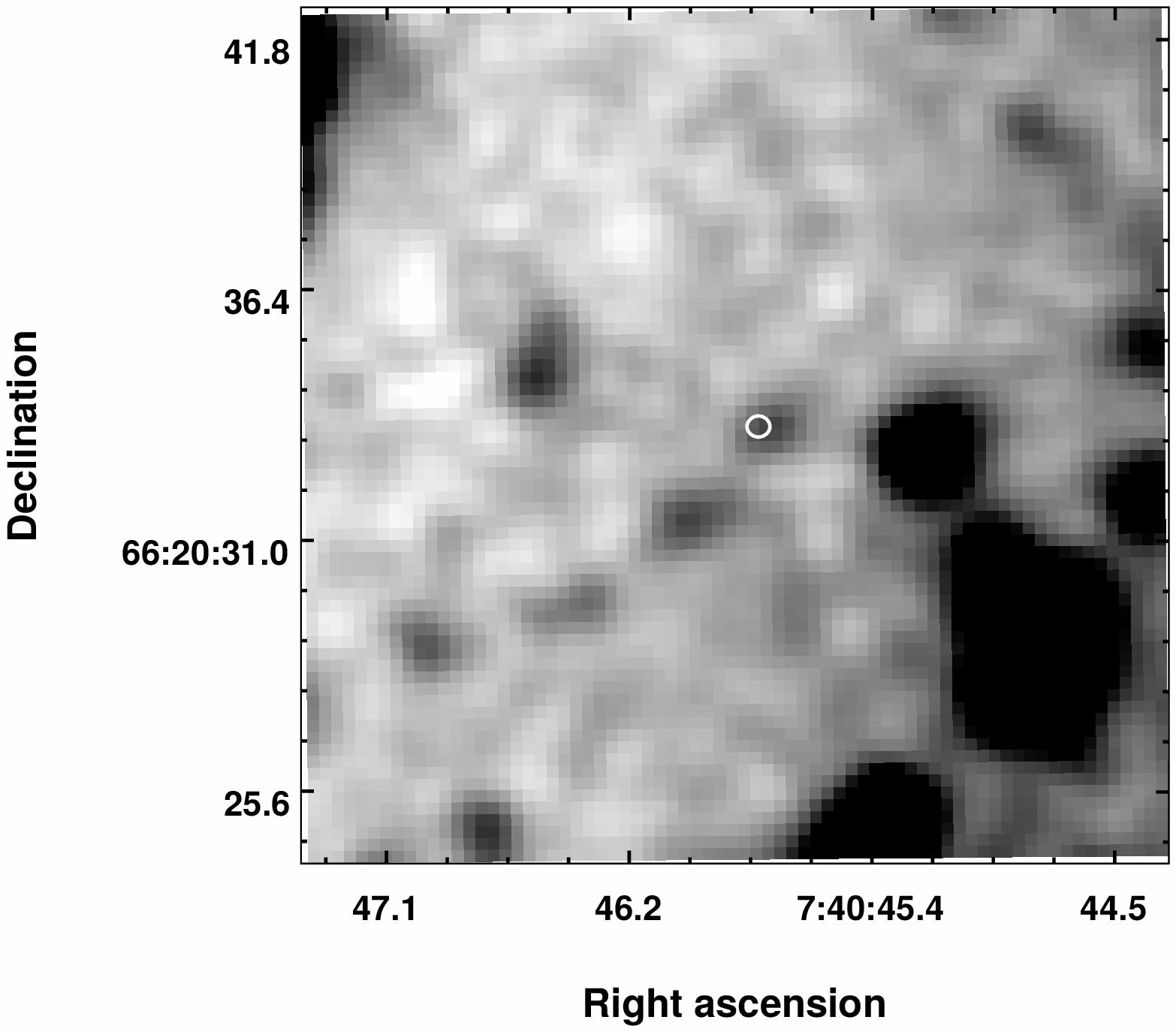}}
\put(65,0){\includegraphics[width=8cm,clip=]{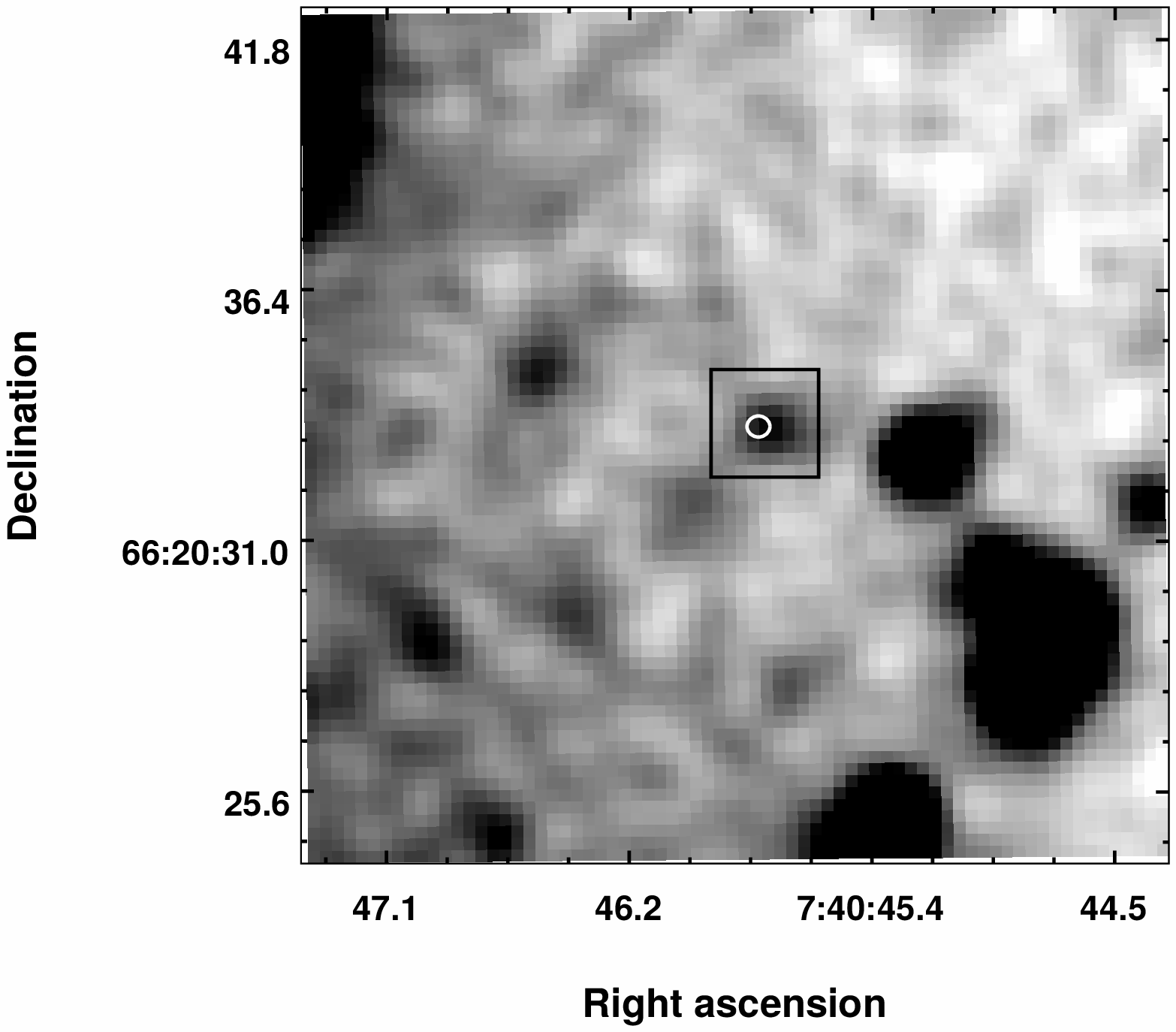}}
\put (119.7,41.5) {\includegraphics[scale=0.11, 
viewport = 0 0 563 653, clip]{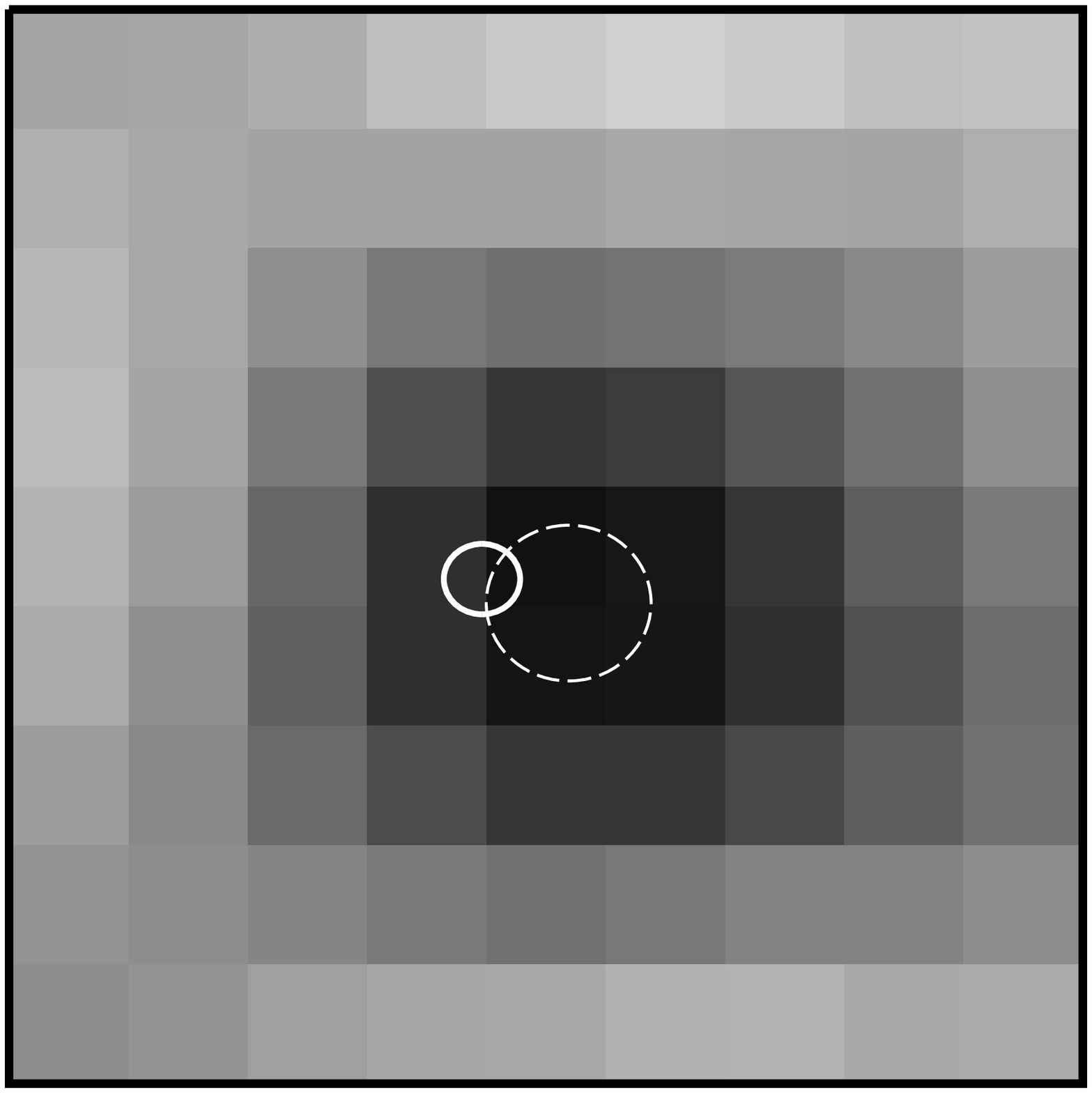}}
\end{picture}
}
\caption{16 arcsec $\times$ 16 arcsec GTC/OSIRIS Sloan $r'$ (left) and $i'$-band (right) images of the \psr\ field. 
The images are smoothed with the Gaussian kernel of 3 pixels, compatible with the seeing value. 
The white circle shows the 3$\sigma$ pulsar position uncertainty of $\approx 0.24$ arcsec, which accounts 
for the optical astrometric referencing errors and the radio timing position uncertainties, corrected for the proper motion. The region of $2.3\times2.3$ arcsec within the black box in the right panel is enlarged 
and shown in the inset at the right-top corner of the image. The solid and dashed circles are 
1$\sigma$ position uncertainties of the pulsar and its possible binary companion, respectively. }
\label{fig:gtc}
\end{figure*}

To perform the astrometric transformation, we used the {\sc iraf ccmap} task 
and 9  bright non-saturated reference stars 
 detected with a high signal-to-noise ratio in the $r'$-band short-exposure frame. 
The best coordinates of the stars are contained in the Gaia DR2 catalogue \citep[][]{gaia2016,gaia2018b} and have 
the catalogue conservative accuracy of $\approx0.7$ mas \citep{lindegren2018}. 
The resulting $1\sigma$ {\it rms} uncertainties of the astrometric fit are  0.08 arcsec in RA and Dec. 
The Gaia catalogue uncertainties are substantially smaller and can be neglected. 
The resulting WCS solution was applied to the combined images in the $r'$ and $i'$ bands. 
Selecting other reference stars did not change the fit result. 

For the photometric calibration, we used the Landolt standard star G163-50 \citep{landolt1992} 
observed during the same night as the target. 
The derived photometric zeropoints for the CCD2 chip are $z_{r'}=28.88\pm0.01$ and $z_{i'}=28.52\pm0.02$. 
The values were obtained by comparing the standard star magnitudes from \citet{smith2002} with 
its instrumental magnitudes, corrected for the finite aperture and atmospheric extinction. 
We used the atmospheric extinction coefficients\footnote{http://www.iac.es/adjuntos/cups/CUps2014-3.pdf} 
$k_{r'}=0.07\pm0.01$ and  $k_{i'}=0.04\pm0.01$.  
The 3$\sigma$ detection limits on the resulting images are $r'<27.5$ and $i'<26.5$.

\section{Results}
\label{sec:results}

Figure~\ref{fig:gtc} shows the sections of the $r'$ and $i'$ images, containing the \psr\ position. 
The radio coordinates of the pulsar corrected for the proper motion (Table \ref{tab:param}) to the epoch 
of our observations (58114 MJD) are RA=07\h40\m45\fss7898(1) and Dec=+66\degs20\amin33\farcs4661(6). 
Accounting for the astrometric referencing uncertainty, the radius of the  \psr\ 1$\sigma$ position error
circle on the optical images is $\approx 0.08$ arcsec.  
In the main panels of Figure~\ref{fig:gtc} we show the 3$\sigma$ circle for convenience. 
A faint starlike source is detected in both bands, which overlaps with the circle. 
Its coordinates RA=07\h40\m45\fss76(1) and Dec=+66\degs20\amin33\farcs4(2) 
were derived as a mean of the source positions in both bands. 
The errors account for the position uncertainties of the object in the frames, 
the difference in the positions in the two bands, and the astrometric 
referencing uncertainty. The object coordinates coincide with the pulsar position  
at the $1\sigma$ significance level. 
This is demonstrated in the inset of the right panel of Figure~\ref{fig:gtc} where
the $1\sigma$ position circles for the pulsar and the object overlap. 

 For photometry of this faint object, we used the circular source aperture 
with $\approx$ 0.76 arcsec (3 pixels) radius, which is consistent with the seeing value 
and provides the maximum signal-to-noise ratio on the source. 
To minimise  possible contributions from wings of nearby sources, backgrounds were 
taken from a narrow  annulus centred on the source 
with the inner and outer radii of $\approx$ 0.8 and 1.5 arcsec. 
As a result, we obtained the source magnitudes 
$r'=26.51\pm0.17$ and $i'=25.49\pm0.15$ corrected for the finite aperture and 
the atmospheric extinction.  
Alternatively, for both bands we constructed the point spread functions (PSFs) 
using about a dozen of 
unsaturated isolated field stars and the {\sc daophot} package, and 
performed the PSF-fitting photometry of field sources applying the {\sc allstar} routine 
with the same source aperture but a wider background annulus 
with the inner and outer radii of $\approx$ 2.5 and 5 arcsec\footnote {The latter is appropriate as the routine iteratively   
subtracts brighter neighbors  in a source group before the photometry of fainter ones.}. 
This yielded the same object magnitudes with only marginally smaller errors. Our object is perfectly PSF-subtracted without any residuals that could indicate a possible presence 
of  fainter sources hidden in its profile.  
Finally, we performed the aperture photometry  on the images where all neighbors,  
excluding the object of  interest, were PSF-subtracted. 
Independently  of the annulus width this led again to the results consistent  with the above aperture and 
PSF magnitude measurements. We thus adopt the above values as the most conservative ones. 

We estimated the probability of detecting a random object at the pulsar position 
using the Poisson distribution $P=1-exp(-{\sigma}S)$,
where $\sigma$ is the surface number density of objects within the GTC field of view 
and $S$ is the area of the pulsar error ellipse. 
As a result, we found the probability of $\approx2\times10^{-3}$ for an accidental coincidence  
of the pulsar and an unrelated object with a brightness of $>$ 20 mag. 
This indicates the source is a likely optical counterpart to the \psr\ binary companion.


\begin{figure}
\begin{minipage}[h]{1.\linewidth}
\center{\includegraphics[width=1.0\linewidth, clip=]{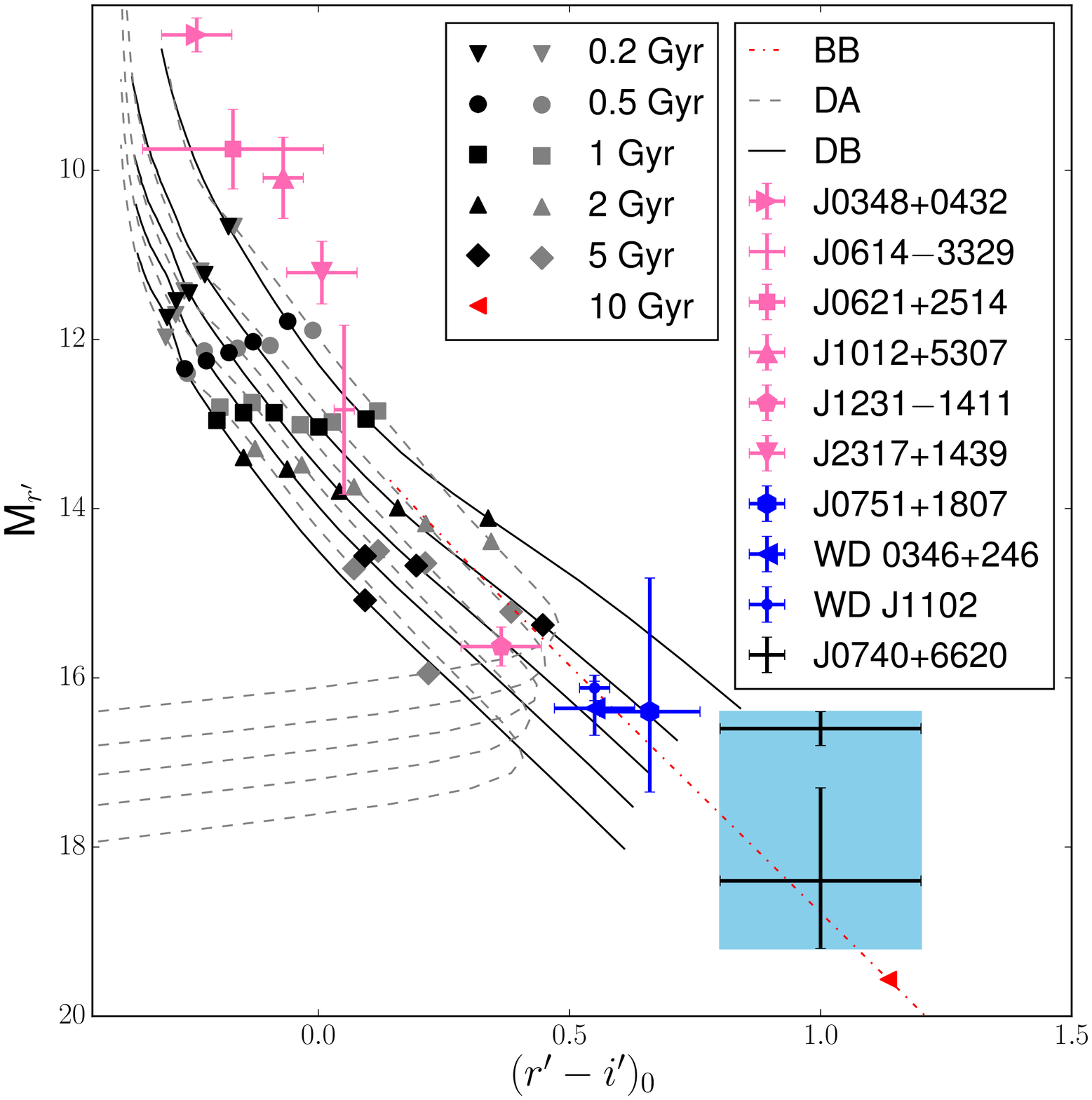}}
\end{minipage}
\caption{Colour-magnitude diagram with various WD evolutionary sequences and the data for different WDs listed in the legend. 
Dashed grey and black solid lines show the cooling tracks for WDs 
with hydrogen (DA) and helium (DB) atmospheres, respectively 
\citep{holberg2006,kowalski2006,tremblay2011,bergeron2011}
with masses 0.2 -- 1.0 \msun~(spaced by 0.2 \msun) increasing from upper to lower curves. 
Cooling ages are indicated by different symbols.
The dash-dotted red line demonstrates the track for a WD with the mass of 0.4 \msun~emitting the blackbody spectrum. 
The location of the \psr\ presumed companion is marked by the black crosses: the upper one ($M_{r'}=16.6^{+0.2}_{-0.2}$)
is for the maximum distance estimate $D_{\rm YMW}=0.93$ kpc and the lower one ($M_{r'}=18.4^{+0.8}_{-1.1}$) is for
the minimum distance estimate $D_p=0.4^{+0.2}_{-0.1}$ kpc, derived from the timing parallax. 
The light-blue rectangle encompasses the   whole distance uncertainty range. 
The data points for WDs that likely have pure helium or mixed atmospheres are shown in blue.}
\label{fig:col-mag}
\end{figure}

\begin{figure}
\begin{minipage}[h]{1.\linewidth}
\center{\includegraphics[width=1.0\linewidth,clip]{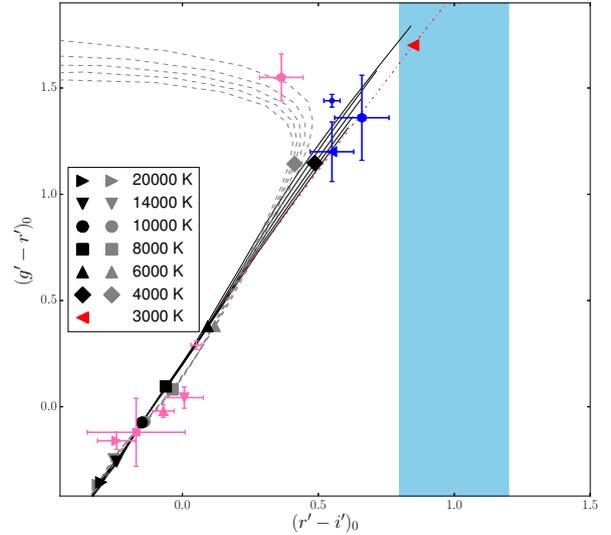}}
\end{minipage}
\caption{Colour-colour diagram with various WD evolutionary sequences and the data for the same WDs as in Fig.~\ref{fig:col-mag}. 
The model predictions presented in Fig.~\ref{fig:col-mag} are 
labelled by the same symbols and colours. 
WD temperatures are indicated by different symbols. 
The location range of the likely \psr\ companion is shown by the light-blue stripe.}
\label{fig:col-col}
\end{figure}

To obtain intrinsic colours and magnitudes of the presumed companion 
we used the empirical correlation between the
distance $D$ and the interstellar reddening $E(B-V)$  
based on the Pan-STARRS~1 and 2MASS photometric data  \citep{dustmap2018}. 
The expected $E(B-V)$ in the pulsar direction is $\approx0.04\pm0.02$ 
for the most conservative distance range of 0.3 -- 0.9 kpc following from Table~\ref{tab:param}. 
This value was transformed to the extinction correction values $A_{r'}=0.09\pm0.05$ and $A_{i'}=0.07\pm0.03$ 
using the conversion coefficients presented by \citet{schlafly2011}. 
The resulting intrinsic magnitudes and colour index are 
$r'_0=26.42\pm0.18$, $i'_0=25.42\pm0.15$, and  $(r'-i')_0=1.0\pm0.2$, 
where the errors include uncertainties of the reddening and magnitude measurements. 
For the above distance range, the absolute magnitude $M_{r'}$ varies from 16.4 to 19.2. 
The specific values for the timing parallax distance  and  for the 
maximum model depended DM distance  of $930$ pc are  $M_{r'}=18.4^{+0.8}_{-1.1}$ and $16.6^{+0.2}_{-0.2}$, respectively. 

As it has been suggested, the \psr\ companion is a WD \citep{lynch2018}. 
Therefore, we compared the derived colour and magnitudes of the likely companion
with the cooling sequences of  hydrogen (DA) and helium (DB) white dwarfs. 
The corresponding colour-magnitude and colour-colour diagrams are presented 
in Figures~\ref{fig:col-mag} and \ref{fig:col-col}, respectively. 
Cooling tracks  for WDs with hydrogen and helium atmospheres 
\citep[known as Bergeron models:][]{holberg2006,kowalski2006,tremblay2011,bergeron2011} are 
shown by different line types\footnote{http://www.astro.umontreal.ca/$\sim$bergeron/CoolingModels}. 
The data points of companions to some other MSPs are presented:
J0348+0432 \citep{antoniadis2013},  J0614$-$3329 \citep{bassa2016}, J0621+2514 \citep{karpova2018}, J0751+1807 \citep{bassa2006}, J1012+5307 \citep{nicastro1995},  J1231$-$1411 \citep{bassa2016} and J2317+1439 \citep{dai2017}. 
We also include  two ultracool isolated WDs: SDSS J110217.48+411315.4 
\citep[hereafter WD J1102;][]{hall2008,kilic2012} and WD 0346+246 \citep{oppenheimer2001}. 
For WD 0346+246 and the PSR J0751+1807 companion, the $UBVRI$ magnitudes were
converted to the Sloan Digital Sky Survey (SDSS) system using transformations by 
Lupton\footnote{https://www.sdss.org/dr14/algorithms/sdssUBVRITransform/}. 
The magnitudes of other objects were adopted either from the
cited articles or the SDSS catalogue \citep{abolfathi2018}. 
The corresponding values of interstellar reddening $E(B-V)$ of the sources were obtained using the models by
\citet{drimmel2003} and \citet{dustmap2018} together with the distances from
DM and parallax measurements or model-predicted WD distances. 
For PSR J2317+1439, the parallax information was updated by \citet{arzoumanian2018},
and for this reason the reddening and absolute magnitude from \citet{dai2017} were recalculated. 
For PSR J0621+2514, we combined the results for two DM distance estimates \citep{karpova2018}. 

The \psr\ presumed companion is the reddest among sources 
shown in the diagrams (Figures~\ref{fig:col-mag},\ref{fig:col-col}). 
 The Bergeron models presented here are constructed for CO-core WDs. 
However, it is generally believed that most of the MSP companions are He-core WDs, due to stripping 
a companion envelope during the mass-transfer binary phase 
\citep[see e.g.][]{rivera-sandoval2015,cadelano2015}. 
For hydrogen-atmosphere WDs, the comparison of the Bergeron models with those of 
the He-core ones created by \citet{althaus2013} shows that the presumed \psr\ companion 
 cannot certainly belong to DA WDs, considering any core composition. 
This is because the upper limit for  $(r'-i')_0$  indices 
of the hydrogen atmosphere sequences is $\la0.5$ for both core types 
and all DA tracks turn towards blue colours at high ($M_{r'}\gtrsim16$) 
absolute magnitudes/low temperatures \citep[see also][]{bassa2016}. At the same time,  
the colour index of the presumed companion is about one (Figure~\ref{fig:col-mag}). 
On the other hand, for cool WDs with pure helium atmospheres colours continue to redden 
with decreasing temperatures. 
Unfortunately, we have not found respective He-core magnitude sequences for these WDs. 
Nevertheless, calculations show that 
the luminosity of a He-core WD at a given age has to be  significantly higher than that of 
the CO-core WD with the same mass \citep[e.g.,][]{oirschot2014}. This is due to the fact that the global specific 
heat capacity of the former is larger than that of the latter. 
However, at ages $\ga 1$~Gyr  the luminosity difference for low mass WDs becomes less than a half of magnitude 
which could not be resolved within a factor of four larger intrinsic magnitude uncertainty of the presumed 
companion. This is mainly caused by the large distance uncertainty (light-blue rectangle in Figure~\ref{fig:col-mag}). 
Moreover, the spectral energy distributions of  cool helium-atmosphere WDs 
become close to those of blackbodies (BBs) \citep[e.g.][]{kilic2011hoard}. 
As an example, in the diagrams we present the cooling sequence for a 0.4 \msun\ WD with the BB spectrum, 
describing the cooling from 6000 to 2000 K. 
One can see that the presumed \psr\ companion position in the diagrams is in agreement with this track. 
Therefore, it most likely belongs to the class of ultracool WDs with helium atmospheres
and has a temperature of $\lesssim3500$~K and a cooling age of $\gtrsim5$~Gyr.

\section{Discussion and conclusions}
\label{sec:discussion}

We detected the likely companion to the MSP \psr. 
The comparison of the photometry results 
with the WD evolutionary sequences showed that the source  
is  most probably an ultracool WD with a pure helium atmosphere. 
It can be the reddest WD among the known MSP  companions. The WD-core composition remains 
unclear due to the large  distance and magnitude uncertainties. 

The presumed WD cooling age is $\gtrsim 5$ Gyr. 
 If it has a low mass, this age limit can be larger
since after the Roche-Lobe detachment
a proto-WD goes through the contraction phase until it reaches 
its cooling track \citep{istrate2014,istrate2016}. 
The duration of this phase increases as the mass of the proto-WD decreases,  
and may last as long as $\sim$ 2 Gyr. 
The \psr\ characteristic age of $3.75$ Gyr (Table~\ref{tab:param}) is smaller than the WD age estimate. 
However, the observed pulsar period derivative and consequently its characteristic age
can be biased by kinematic effects, i.e. the effects of 
the pulsar proper motion \citep[Shklovskii effect;][]{shklovskii1970},
the acceleration towards the Galactic plane and the acceleration due to differential
Galactic rotation \citep{nice1995}. 
 Using the \psr\ proper motion value from Table~\ref{tab:param},
the Sun's galactocentric velocity and the distance \citep[240 km~s$^{-1}$ and 8.34 kpc, respectively;][]{reid2014}, 
we calculated these corrections to the pulsar period derivative:
$\dot{P}_{\rm S}=3.0\times10^{-21}$,
$\dot{P}_{\rm G,\perp}=-1.6\times10^{-22}$,
$\dot{P}_{\rm G,p}=3.8\times10^{-23}$ for the minimum and
$\dot{P}_{\rm S}=6.9\times10^{-21}$,
$\dot{P}_{\rm G,\perp}=-2.2\times10^{-22}$,
$\dot{P}_{\rm G,p}=8.9\times10^{-23}$
for the maximum pulsar distance estimates.
The corresponding intrinsic characteristic ages are $\tau_i\sim5$ Gyr 
and $\sim 8.5$ Gyr, which are compatible with the cooling age\footnote{ 
It is well known that the characteristic age
is just an estimate of the pulsar true age and can significantly deviate from it
\citep[e.g.][]{psrhandbook}. However, it is usually the only available estimate 
which is the case of \psr. The pulsar true age can be even larger 
than $\tau_i$ if the braking index is $<$ 3.}.
Thus, the considered binary system indeed can be very old and the presumed WD can be ultracool.
This is not a unique situation. There are other examples of the objects with similar characteristics.
One of them
is the WD companion of PSR J0751+1807  \citep{bassa2006}.
Its colours (see Figure~\ref{fig:col-mag} and \ref{fig:col-col}) indicate that the WD has a pure helium or mixed H/He atmosphere with a temperature $T\sim$ 3500 -- 4300 K. 
Other examples are isolated ultracool white dwarfs WD J1102 
\citep{hall2008,kilic2012} and WD 0346+246 \citep{oppenheimer2001}.
These WDs have temperatures of about 3650 and 3300 K, 
respectively, and are best explained by the mixed atmosphere models
\citep[][]{gianninas2015}. 

\begin{figure}
\begin{minipage}[h]{1.\linewidth}
\center{\includegraphics[width=0.9\linewidth,clip]{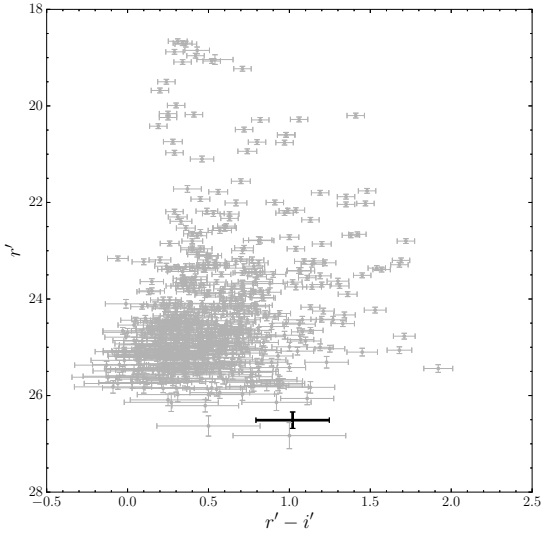}}
\end{minipage}
\caption{Observed colour-magnitude diagram for about six hundred stellar sources in the \psr\ field based on the GTC observations. 
The likely pulsar companion is shown in black.}
\label{fig:col-mag_field}
\end{figure}

It is clear that a single colour index is not enough 
to confidently assert that the source is the optical counterpart 
of the pulsar companion. 
Although the probability of the accidental coincidence is low, 
however, it cannot be excluded. 
 As seen from the colour-magnitude diagram presented in Figure~\ref{fig:col-mag_field},  
the presumed companion is not very distinct from the distribution of other stellar sources  
detected within the GTC FOV. 
It is located near the red locus of the main sequence distribution. 
Below we consider possible alternative interpretations of the source nature.

Firstly, the source colour is also in agreement with that of red/brown dwarfs
which are cool low-mass ($M\sim$ 0.08 -- 0.6 \msun) main-sequence stars. 
According to the correlation between the colour and the spectral type,  
based on the SDSS sample of M, L, and T dwarfs \citep{hawley2002},
the source can be a M0--M3V red dwarf  or a T-type brown dwarf. 
The corresponding distance of a red dwarf is $\gtrsim 8$ kpc, and the object with $b=+29.6$ deg (see Table~\ref{tab:param}) must be located at $\gtrsim$~4~kpc above the Galactic disk. A T-type brown dwarf would be very close to the Sun at a distance $\lesssim50$ pc. 
Both values are not consistent with the distance 
to \psr~and, in this case, the source must be rejected as  
the MSP companion.

Another possibility is that the source is a red subdwarf.
Red subdwarfs are the metal-poor analogues of late-type dwarfs
\citep[see e.g.][]{savcheva2014, zhang2013}. 
Absolute magnitudes of most of the detected late-type subdwarfs are
$9\lesssim M_{r'}\lesssim14$, which infer a distance of $\gtrsim3$ kpc to the source. This is also incompatible with the distance to \psr\ and implies the object is outside the Galactic disk.

Finally,  
the extragalactic origin of the source cannot be excluded. 
The total Galactic extinction in this direction $E_{B-V}\lesssim0.06$ is relatively low and such colour is expected, e.g., for distant E/S0/Sbc-type galaxies with redshifts $z\leq1$ \citep[see e.g.][]{georgakakis2006}. 

If the source is not related to the \psr, 
we can consider the derived magnitudes as lower limits for the pulsar
binary companion. 
In this case it could be even cooler and older though we cannot determine the type of its atmosphere. 
Deep infrared observations are necessary to clarify the properties and nature of the detected source. For a cool WD with a pure helium atmosphere one can expect the blackbody-like spectral energy distribution. Future proper motion measurements can confirm the association of the likely counterpart with \psr.

\section*{Acknowledgements}

 We thank the anonymous referee for useful comments and
P. Bergeron for helpful discussions. The work of DMB, AVK and DAZ (Sec.~\ref{sec:data} and \ref{sec:results}) was supported by the Russian Foundation for Basic Research, 
project No. 18-32-00781 mol\_a. AYuK and SVZ acknowledge PAPIIT grant IN-100617 for resources provided towards this research. The work is based on observations made with the Gran Telescopio Canarias (GTC), installed at the Spanish Observatorio del Roque de los Muchachos of the Instituto de Astrof\'isica de Canarias, in the island of La Palma. {\sc iraf} is distributed by the National Optical Astronomy Observatory, which is operated by the Association of Universities for Research in Astronomy (AURA) under a cooperative agreement with the National Science Foundation. 
{This work has made use of data from the European Space Agency (ESA) mission {\it Gaia} (\url{https://www.cosmos.esa.int/gaia}), processed by the {\it Gaia}
Data Processing and Analysis Consortium (DPAC, \url{https://www.cosmos.esa.int/web/gaia/dpac/consortium}). Funding for the DPAC has been provided by national institutions, in particular the institutions
participating in the {\it Gaia} Multilateral Agreement. DAZ thanks Pirinem School
of Theoretical Physics for hospitality.}



\bibliographystyle{mnras}

\bibliography{msp4} 

\bsp	
\label{lastpage}
\end{document}